\documentclass[aps,showpacs,superscriptaddress,twocolumn]{revtex4}%

\usepackage{amsfonts}
\usepackage{amsmath}
\usepackage{amssymb}
\usepackage{graphicx}%
\begin{document}
\title{Topological phases and fractional excitations of exciton condensate in a
special class of bilayer systems}
\author{Ningning Hao}
\affiliation{Institute of Physics, Chinese Academy of Sciences, Beijing 100190, P. R. China.}
\affiliation{LCP, Institute of Applied Physics and Computational Mathematics, P.O. Box
8009, Beijing 100088, People's Republic of China}
\author{Ping Zhang}
\thanks{Corresponding author. zhang\_ping@iapcm.ac.cn}
\affiliation{LCP, Institute of Applied Physics and Computational Mathematics, P.O. Box
8009, Beijing 100088, People's Republic of China}
\affiliation{Center for Applied Physics and Technology, Peking University, Beijing 100871,
People's Republic of China}
\author{Yupeng Wang}
\affiliation{Institute of Physics, Chinese Academy of Sciences, Beijing 100190, P. R. China.}


\begin{abstract}
We study the exciton condensate in zero temperature limit in a special class
of electron-hole bilayer systems adjacent to insulating ferromagnetic films.
With the self-consistent mean-field approximation, we find that the Rashba
spin-orbit interaction in the electron and hole layers can induce the $p\pm
ip$ or $p$ pairing states depending on the different magnetization of the
overlapped ferromagnetic films. Correspondingly, the topologically nontrivial
or trivial phases emerge. Furthermore, in the topologically nontrivial phase,
the quasiparticle excitations of the $U(1)$ vortex are attached to fractional
quantum numbers and obey Abelian statistics.

\end{abstract}
\maketitle

\section{Introduction}

Most phases in condensed matter physics can be well understood with Landau's
phase transition theory, which characterizes states of matter in terms of
local order parameters associated with spontaneous breaking of the underlying
symmetries. However, the quantum Hall (QH) state \cite{Klitzing,Tsui} gives
the first example of topological order \cite{Thouless,Wen} without breaking
any symmetries and can be characterized by topological quantum numbers and
gapless edge states. From then on, search for possible realization of
nontrivial topological phases has become an intriguing and challenging issue
in basic condensed matter physics as well as technological applications.
Recently, the two dimensional (2D) topological band insulators (TBI)
\cite{Haldane,Kane1,Kane2,Zhang} present new topological phases which have QH
effect or quantum spin Hall (QSH) effect depending on time-reversal symmetry
(TRS) broken or not. Correspondingly, The 2D TBI can be well characterized by
the (Thouless, Kohmoto, Nightingale, and Nijs) TKNN number \cite{Thouless} for
the QH phases and by $Z_{2}$ topological number \cite{Kane2} for the QSH
phases. Soon after, the 3D TBIs, which are the natural generalization from 2D
TBIs, were discovered in several real materials such as Bi$_{1-x}$Sb$_{x}$
alloys and Be$_{2}$Se$_{3}$-family crystals
\cite{Fu2007,Hsieh,Zhang2009,Xia2009,Chen2009}.

Besides 2D and 3D TBIs, more recently, a new class of topological
superconductors (TSs) has been predicted by the topological classification of
the Bogoliubov-de-Gennes (BdG) Hamiltonians \cite{Schynder,Kitaev}. Most
interestingly, the topological quasiparticle excitations in TSs are Majorana
Fermions with non-Abelian statistics \cite{Ivanov,Read}, which have potential
applications in topological quantum computation \cite{Nayak}. In analogy with
TS, recently, we gave a proposal to realize the topological exciton condensate
(TEC) in a spin-orbit coupled electron-hole bilayer system adjacent to two
insulating ferromagnetic (FM) films \cite{Hao}. Note that in our previous
proposal \cite{Hao}, the electron-hole bilayer is fabricated with
semiconductors such as GaAs or InAs heterostructures, where the electron and
hole have different effective mass and different type of Rashba spin-orbit
coupling (namely, $k$-type in electron layer and $k^{3}$-type in hole layer).
Unfortunately, these differences break the particle-hole symmetry (PHS) of the
bilayer system considered in Ref. \cite{Hao}. We know that the PHS is
indispensable for emergence of the stably topological quasiparticle
excitations, such as Majorana fermion in TSs \cite{Sau,Sato}. Hence, we can
predict no stable topological quasiparticle excitations exist in TEC proposed
in Ref. \cite{Hao}. In this paper, we extend the work of Ref. \cite{Hao} by
considering the presence of PHS and discuss the topological quasiparticle
excitations. \begin{figure}[ptb]
\begin{center}
\includegraphics[width=1.0\linewidth]{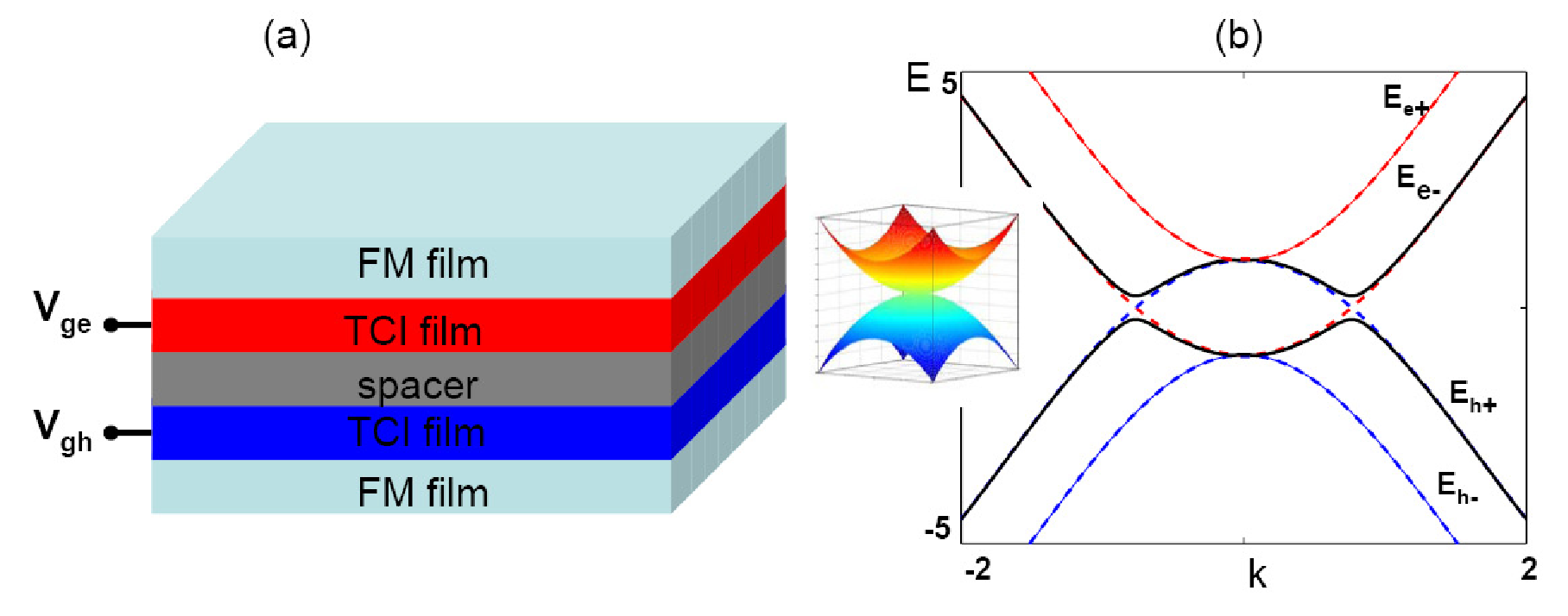}
\end{center}
\caption{(Color online) (a) Schematic structure of the effective electron-hole
bilayer system that holds the exciton condensate. Different layers are
fabricated with corresponding material shown on the layers, in which the
spacer are fabricated with dielectric and its thickness is $d$. The electron
and hole layers are obtained by tuning the gate voltages $V_{ge}$ and $V_{gh}%
$. The inserted picture is the dispersion of the surface state of TCI. The FM
films provide effective exchange fields ($V_{e},V_{h}$). (b) The energy
spectrum of the effective electron-hole bilayer system near the Fermi energy
$E_{F}$=$0$. Here the dashed lines denote non-interacting single-particle
energy spectrum $E_{e(h)\pm}$, while the solid lines denote the exciton energy
spectrum with an obvious mean-field gap opened. We take $t_{e}=t_{h}=1$,
$\mu_{e}=\mu_{h}=-4$, $(V_{e},V_{h})=(-1,-1)$, and $\alpha=0.5$.}%
\end{figure}

We start with an electron-hole bilayer system shown in Fig. 1(a). In order to
preserve the PHS, it is not suitable for the electron and hole layers to be
fabricated with semiconductor heterostructures due to the aforementioned
reasons. We argue that the recently predicted topological crystalline
insulator (TCI) \cite{Fu2011} films are appropriate candidates which have
metallic surface states with quadratic dispersion on high-symmetry crystal
surfaces. The robustness of the surface states of TCI are protected by the
crystal point group symmetries [for example four-fold ($C_{4}$) or six-fold
($C_{6}$) rotational symmetry] associated with TRS of spinless fermion. Hence,
the TCIs are the counterpart of $Z_{2}$ topological insulators without
spin-orbit coupling. Correspondingly, when TRS of spinless fermion is broken,
the TCIs are the counterpart of TKNN topological insulators. When the suitable
parameters are chosen in the TCI model, the dispersion of the surface states
can be described as $\epsilon_{k}=\pm k^{2}/2m_{eff}$. Then the electron and
hole layers can be obtained by independently modulating the gate voltages
attached to the TCI films. Generally, only the orbital degrees of freedom are
considered in TCIs. However, the spin degrees of freedom have to be considered
if the magnetic effect is introduced, which is the case in our model. We
introduce two insulating FM films adjacent to the two TCI films to support two
effective interface exchange fields, which respectively induce the population
imbalance of the spin-up and spin-down carriers in the electron and hole
layers and break the spinful TRS of the system due to the Zeeman effect of the
exchange fields. We argue that the magnetic vector potential (orbital effect)
of the exchange fields of the FM films can be neglected, because the exchange
fields only locate on the interface of the FM films and TCI films. That is the
reason why we do not apply external magnetic fields to replace the FM films.
Additionally, we assume that the exchange fields are always perpendicular to
the bilayer. It is predictable that the interlayer Coulomb interaction induces
the excitonic instability when the two layers are separated with an insulating
spacer and close enough with each other. Moreover, We note that since
structure inversion symmetry is broken at the TCI film and the insulating
spacer interface, there will be an interface Rashba spin-orbit coupling (SOC)
\cite{Rashba,Bychkov}: $\mathcal{H}_{R}=\alpha(\sigma\times\mathbf{k}%
)\cdot\hat{z}$ and $\alpha\propto\left\langle \nabla_{z}V\right\rangle \,$,
where $V$ comes from doping, impurities, external gate voltages and so on. It
is evident to find that $\alpha$ is not zero when the structure inversion
symmetry is broken and can be tunable by external gate voltages. Moreover, a
Dresselhaus term \cite{Dresselhaus} can not be zero when the bulk inversion
symmetry is broken. In the two dimensions, $\mathcal{H}_{D}=\beta(k_{x}%
\sigma_{x}-k_{y}\sigma_{y})$ and $\beta\approx-\mathcal{B}(\pi/w)^{2}$, where
$\mathcal{B}$ is a material-dependent parameter. In the conventional
semiconductor quantum-well systems, $w$ is the width of the well with a
typical value of $\sim$50 \r{A}, then $\beta\mathtt{\sim}0.3$ $(0.1)$ eV \r
{A}\ in InSb (GaAs) quantum wells, which is comparable to or even larger than
the value of $\alpha$ \cite{Jason}. In this case, the Dresselhaus term can not
be neglected. However, if the electron and hole layers are fabricated with TCI
films, since the TCI has the bulk inversion symmetry, then the Dresselhaus
term can be neglected and we can only consider the Rashba term. When the
amplitude of the Rashba SOC is comparable to that of the exciton order
parameter (EOP), we find that the unconventional pairing states emerge with
topologically nontrivial $p$+$ip$ pairing states or trivial $p$ pairing states
depending on the magnetization of the overlapped FM films. Furthermore, we
find that the topological quasiparticle excitations of the $U(1)$ vortex
profiles survive and carry fractional quantum numbers \cite{Jackiw,Su} that
obey the Abilan statistics.

The paper is organized as follows. In Sec. II a model Hamiltonian for the
effective electron-hole bilayer system is introduced, and a self-consistent
mean-field method is used to produce the unconventional pairing states. In
Sec. III we introduce the TKNN number as the topological number to
characterize topologically nontrivial pairing states and show the chiral
gapless edge states. In Sec. IV we analytically study the topological
quasiparticle excitation of a single $U(1)$ vortex in the low-energy limit and
give the numerical results in lattice model and find both are consistent. In
Sec. V we summarize our results.

\section{Model Hamiltonian and Exciton Condensate}

Consider the effective TCI electron-hole bilayer confined in the $x$-$y$ plane
in zero temperature limit. Then the mean-field description is appropriate even
in two dimensions. For convenience of discussion, here we consider a square
lattice model. Furthermore, the continuous model can be obtained from the
low-energy limit of the lattice model. The square lattice Hamiltonian is
$\mathcal{H}$=$\sum_{p}(\mathcal{H}_{kin}^{(p)}$+$\mathcal{H}_{R}^{(p)}%
)$+$\mathcal{H}_{int}^{(e-h)}$:%
\begin{align}
\mathcal{H}_{kin}^{(p)}  &  =\underset{<i,j>,\sigma}{\sum}(-t_{p}-\mu
_{p}\delta_{ij})\hat{p}_{i,\sigma}^{\dagger}\hat{p}_{j,\sigma}\nonumber\\
&  +\underset{j,\sigma,\sigma^{\prime}}{\sum}V_{p}(s_{z})_{\sigma
,\sigma^{\prime}}\hat{p}_{j,\sigma}^{\dagger}\hat{p}_{j,\sigma^{\prime}%
},\nonumber\\
\mathcal{H}_{R}^{(p)}  &  =\frac{1}{2}\alpha_{p}[\underset{j}{\sum}(\hat
{p}_{j,\uparrow}^{\dagger}\hat{p}_{j+\delta x,\downarrow}-\hat{p}_{j,\uparrow
}^{\dagger}\hat{p}_{j-\delta x,\downarrow})\nonumber\\
&  -i\underset{j}{\sum}(\hat{p}_{j,\uparrow}^{\dagger}\hat{p}_{j+\delta
y,\downarrow}-\hat{p}_{j,\uparrow}^{\dagger}\hat{p}_{j-\delta y,\downarrow
})]+\text{H.c.},\nonumber\\
\mathcal{H}_{int}^{(e-h)}  &  =-\underset{i,j,\sigma,\sigma^{\prime}}{%
{\displaystyle\sum}
}U_{i,j}(d)\hat{e}_{i\mathbf{\sigma}}^{\dagger}\hat{h}_{j\mathbf{\sigma
}^{\prime}}^{\dagger}\hat{h}_{j\mathbf{\sigma}^{\prime}}\hat{e}%
_{i\mathbf{\sigma}}\text{.} \label{H_lat}%
\end{align}
Here $t_{p}$ denotes the nearest-neighbor hopping amplitude while $\mu_{p}$
represents the chemical potential in electron ($p$=$e$) or hole ($p$=$h$)
layer. Without loss of generality, we set $t_{p}$=$1$ for both electron and
hole layer, other parameters are measured in $t_{p}$. $s_{x,y,z}$ are the
three Pauli matrices and $s_{0}$ is the identity matrix. $V_{p}$ represents
Zeeman splitting from the effective interface exchange field of the FM film.
$\hat{p}_{j,\sigma}$ is the annihilation operator of carrier with spin
$\sigma$ at lattice site $j$. $\alpha_{e(h)}$ is the Rashba SO interaction
strength in the electron (hole) layer. For simplicity, in this paper we set
$\alpha_{e}$=$\alpha_{h}$=$\alpha$. $\delta x$ ($\delta y$) is the
square-lattice spacing measured in unit of in-plane lattice constant along the
$x$ ($y$) direction. The interlayer Coulomb interaction is $U_{i,j}(d)$%
=$e^{2}/\varepsilon\sqrt{\left\vert \mathbf{r}_{i,e}-\mathbf{r}_{j,h}%
\right\vert ^{2}+d^{2}}$, where $\varepsilon$ is the dielectric constant of
the spacer and $d$ is the interlayer distance. We only consider the
interaction relevant to exciton formation and ignore the electron-hole
exchange interaction. We also neglect the intralayer electron-electron and
hole-hole interactions, since they are expected to renormalize the
single-particle spectrum of each layer and have no essential influence on the
topological properties of the system.

In momentum space, the Hamiltonian Eq. (1) can be expressed as
\begin{align}
\mathcal{H}_{kin}^{(p)}  &  =\underset{\mathbf{k},\sigma,\sigma^{\prime}}{%
{\displaystyle\sum}
}[(\zeta_{\mathbf{k}}^{(p)}-\mu_{p})\delta_{\sigma,\sigma^{\prime}}%
+V_{p}(s_{z})_{\sigma,\sigma^{\prime}}]\hat{p}_{\mathbf{k},\sigma}^{\dagger
}\hat{p}_{\mathbf{k},\sigma^{\prime}},\nonumber\\
\mathcal{H}_{R}^{(p)}  &  =\underset{\mathbf{k}}{%
{\displaystyle\sum}
}i\alpha(\sin k_{x}-i\sin k_{y})\hat{p}_{\mathbf{k}\uparrow}^{\dagger}\hat
{p}_{\mathbf{k}\downarrow}+\text{H.c.},\nonumber\\
\mathcal{H}_{int}^{(e-h)}  &  =\underset{\mathbf{kk}^{\prime}\mathbf{q}%
\sigma\sigma^{\prime}}{%
{\displaystyle\sum}
}-\frac{U(\mathbf{q})}{\Omega}\hat{e}_{\mathbf{k+q\sigma}}^{\dag}\hat
{h}_{\mathbf{k}^{\prime}-\mathbf{q\sigma}^{\prime}}^{\dag}\hat{h}%
_{\mathbf{k}^{\prime}\mathbf{\sigma}^{\prime}}\hat{e}_{\mathbf{k\sigma}%
}\text{,} \label{Ham1_k}%
\end{align}
where $U(\mathbf{q})\mathtt{=}\frac{2\pi e^{2}}{\varepsilon q}e^{-qd}$ and
$\zeta_{\mathbf{k}}^{(p)}$=$-2t_{p}(\cos k_{x}$+$\cos k_{y})$. In the Nambu
notation with combined $\hat{e}$-$\hat{h}$ field operator basis $\hat{\Psi
}_{\mathbf{k}}$=$[\hat{e}_{\mathbf{k\uparrow}}$ $\hat{e}_{\mathbf{k\downarrow
}}$ $\hat{h}_{-\mathbf{k\uparrow}}^{\dagger}$ $\hat{h}_{-\mathbf{k\downarrow}%
}^{\dagger}]^{T}$, the decoupled mean-field Hamiltonian is expressed as
$\mathcal{H}_{MF}$=$\hat{\Psi}^{\dagger}(\mathcal{H}_{0}+\mathcal{H}_{1}%
)\hat{\Psi}$+$E_{0}$ in the following Bogoliubov--de Gennes (BdG) form:
\begin{equation}
\mathcal{H}_{MF}=\sum_{\mathbf{k}}\hat{\Psi}_{\mathbf{k}}^{\dagger}\left[
\begin{array}
[c]{cc}%
\mathbf{\Sigma}_{\mathbf{k}}^{(e)}-\mu_{e}+V_{e}s_{z} & \mathbf{\Delta
}(\mathbf{k})\\
\mathbf{\Delta}^{\dagger}(\mathbf{k}) & \mathbf{\Sigma}_{-\mathbf{k}}%
^{(h)}+\mu_{h}-V_{h}s_{z}%
\end{array}
\right]  \hat{\Psi}_{\mathbf{k}}, \label{Ham1_mf1}%
\end{equation}
where $\mathcal{H}_{0}$ represents the single-particle terms while
$\mathcal{H}_{1}$ includes the interaction terms, $\mathbf{\Sigma}%
_{\pm\mathbf{k}}^{(p)}$=$\pm\zeta_{\pm\mathbf{k}}^{(p)}s_{0}\mathtt{\pm
}\mathcal{H}_{R}^{(p)}$($\pm\mathbf{k}$), $\Delta_{\sigma\sigma^{\prime}%
}(\mathbf{k})$=$-\frac{1}{\Omega}\sum_{\mathbf{q}}U(\mathbf{q})\left\langle
\hat{h}_{-\mathbf{k}+\mathbf{q\sigma}^{\prime}}\hat{e}_{\mathbf{k}%
-\mathbf{q\sigma}}\right\rangle $ are the EOPs, and $E_{0}$=$\frac{1}{\Omega
}\sum_{\mathbf{k,q}\sigma\sigma^{\prime}}\frac{\Delta_{\sigma\sigma^{\prime}%
}(\mathbf{k})\Delta_{\sigma\sigma^{\prime}}^{\ast}(\mathbf{k}-\mathbf{q}%
)}{U(\mathbf{q})}$ in which we have neglected the inessential constants.

Moreover, the exciton condensate with unconventional pairing can be well
understood in the Fermi surface nesting picture. To reveal this fact, firstly,
the non-interacting Hamiltonian $\mathcal{H}_{0}$ in Eq. (\ref{Ham1_mf1}) can
be diagonalized with the unitary transformation in the single-particle
eigenstate space as%

\begin{equation}
\mathcal{H}_{0}=%
{\displaystyle\sum_{\mathbf{k}}}
\hat{\Psi}_{0\mathbf{k}}^{\dag}\left(  H_{E}^{(e)}(\mathbf{k})\oplus
H_{E}^{(h)}(\mathbf{k})\right)  \hat{\Psi}_{0\mathbf{k}},\label{H_single}%
\end{equation}
where $H_{E}^{(p)}(\mathbf{k})$=$E_{p-}(\mathbf{k})(s_{0}+s_{z})/2$%
+$E_{p+}(\mathbf{k})(s_{0}\mathtt{-}s_{z})/2$, $\hat{\Psi}_{0\mathbf{k}}%
$=$[\hat{\psi}_{e+}(\mathbf{k}),\hat{\psi}_{e-}(\mathbf{k}),\hat{\psi}%
_{h+}^{\dag}(\mathbf{k}),\hat{\psi}_{h-}^{\dag}(\mathbf{k})]^{T},$ $E_{ps}%
(k)$=$\pm\zeta_{\pm\mathbf{k}}^{(p)}\mathtt{\mp}\mu_{p}\mathtt{+}s\sqrt
{\alpha^{2}(\sin^{2}k_{x}\text{+}\sin^{2}k_{y})\text{+}V_{p}^{2}}$ ($s$=$+,-$)
are respectively electron and hole band energies, and $\hat{\psi}_{ps}$ denote
the relevant particle annihilation operators. The two basis sets are related
by the corresponding unitary transformation, $\hat{\Psi}_{\mathbf{k}}%
$=$e^{i\theta_{\mathbf{k}}}\Pi_{e}\oplus e^{i\vartheta_{\mathbf{k}}}\Pi
_{h}^{\ast}$ $\hat{\Psi}_{0\mathbf{k}}$ with

\begin{equation}
\Pi_{p}=\left[
\begin{array}
[c]{cc}%
-if_{p+}(k)e^{-i\phi_{\mathbf{k}}} & f_{p-}(k)\\
f_{p-}(k) & -if_{p+}(k)e^{i\phi_{\mathbf{k}}}%
\end{array}
\right]  , \label{unitary}%
\end{equation}
where $\phi_{\mathbf{k}}$=$\arctan(\sin k_{y}/\sin k_{x})$ and $f_{p\pm
}(\mathbf{k})$=$\omega_{\mathbf{k}}/\sqrt{\omega_{\mathbf{k}}^{2}%
+(\sqrt{\omega_{\mathbf{k}}^{2}+V_{p}^{2}}\pm V_{p})^{2}}$ with $\omega
_{\mathbf{k}}$=$\alpha\sqrt{\sin^{2}k_{x}+\sin^{2}k_{y}}$. It should be
stressed that the $k$-dependent phases $\theta_{k}$ and $\vartheta_{k}$ are
confirmed by exactly solving the ground state of the system through
self-consistent mean-field calculation. The single-particle bands $E_{p\pm
}(\mathbf{k})$ are shown in Fig. 1(b) (dashed curves), in which the occupied
bands are $E_{e-}$ and $E_{h+}$. With the values of the tunable parameters
shown in the caption in Fig. 1, the band $E_{e-}$ and band $E_{h+}$ have the
perfect identical Fermi surface. From the nesting mechanism, the attractive
interlayer interaction leads to unstability of the Fermi surfaces and opens a
gap. This is the case similar to the conventional BCS picture. Now, the
electron-hole interaction part relevant to exciton formation in Eq.
(\ref{Ham1_k}) can be expressed in terms of the occupied electron band
$E_{e-}$ and hole band $E_{h+}$ with Eq. (\ref{unitary}). Then, the decoupled
mean-field two-band Hamiltonian of the exciton system under the basis
$\bar{\Psi}$=$\left[  \hat{\psi}_{e-}(\mathbf{k}),\hat{\psi}_{h+}^{\dagger
}(-\mathbf{k})\right]  ^{T}$reads (with inessential constants neglected)%
\begin{equation}
\mathcal{\bar{H}}_{MF}=\sum_{\mathbf{k}}\bar{\Psi}_{\mathbf{k}}^{\dagger
}\left[
\begin{array}
[c]{cc}%
E_{e-} & \bar{\Delta}(\mathbf{k})\\
\bar{\Delta}^{\ast}(\mathbf{k}) & E_{h+}%
\end{array}
\right]  \bar{\Psi}_{\mathbf{k}}+\bar{E}_{0}, \label{H_mf2}%
\end{equation}
where $\bar{\Delta}(\mathbf{k})$=$-\frac{1}{\Omega}\underset{\mathbf{q,}%
s,s^{\prime}}{%
{\displaystyle\sum}
}U_{re}(\mathbf{k},\mathbf{q},s,s^{\prime})\left\langle \psi_{h+}%
(-\mathbf{k}\text{+}\mathbf{q})\psi_{e-}(\mathbf{k}\text{-}\mathbf{q}%
)\right\rangle $ and $\bar{E}_{0}$=$\frac{1}{\Omega}\underset{\mathbf{kq}}{%
{\displaystyle\sum}
}\frac{\bar{\Delta}(\mathbf{k})\bar{\Delta}^{\ast}(\mathbf{k}^{\prime}%
)}{\underset{ss^{\prime}}{%
{\displaystyle\sum}
}U_{re}^{\ast}(\mathbf{k},\mathbf{q},s,s^{\prime})}$ with renormalized
interaction $U_{re}(\mathbf{k},\mathbf{q},s,s^{\prime})\mathbf{=}%
U(\mathbf{q})\chi_{\mathbf{k},\mathbf{q}}F_{ss^{\prime}}(\mathbf{k}%
,\mathbf{q})\tau_{\mathbf{k},\mathbf{q}}$, $s$=$\pm$, $\chi_{\mathbf{k}%
,\mathbf{q}}$=$e^{i(\theta_{\mathbf{k}-\mathbf{q}}-\theta_{\mathbf{k}}%
)}e^{i(\vartheta_{\mathbf{k}}-\vartheta_{\mathbf{k}-\mathbf{q}})}$, and
$F_{ss^{\prime}}(\mathbf{k},\mathbf{q})\mathbf{=}f_{es}(k)f_{es}%
(k\mathtt{-}q)f_{es^{\prime}}(k)f_{es^{\prime}}(k\mathtt{-}q)$. Here,
$\tau_{\mathbf{k},\mathbf{q}}$=$e^{i(\phi_{\mathbf{k}-\mathbf{q}}%
-\phi_{\mathbf{k}})}$ for $s\mathtt{\neq}s^{\prime},$ $\tau_{\mathbf{k}%
,\mathbf{q}}$=$1$ for $s$=$s^{\prime}$=$-,$ and $\tau_{\mathbf{k},\mathbf{q}}%
$=$e^{i2(\phi_{\mathbf{k}-\mathbf{q}}-\phi_{\mathbf{k}})}$ for $s$=$s^{\prime
}$=$+$. Then the EOPs in Eq. (\ref{Ham1_mf1}) and those in Eq. (\ref{H_mf2})
are related by the unitary transformation associated with Eq. (\ref{unitary}), i.e.,%

\begin{align}
\mathbf{\Delta}(\mathbf{k})  &  =-e^{i(\theta_{\mathbf{k}}-\vartheta
_{\mathbf{k}})}\bar{\Delta}(\mathbf{k})\nonumber\\
&  \times\left[
\begin{array}
[c]{cc}%
if_{e+}(\mathbf{k})f_{h+}e^{-i\phi_{\mathbf{k}}} & f_{e+}(\mathbf{k}%
)f_{h-}(\mathbf{k})\\
-f_{e-}(\mathbf{k})f_{h+}(\mathbf{k}) & if_{e-}(\mathbf{k})f_{h-}%
(\mathbf{k})e^{i\phi_{\mathbf{k}}}%
\end{array}
\right]  . \label{relation}%
\end{align}

\begin{figure}[ptb]
\begin{center}
\includegraphics[width=1.0\linewidth]{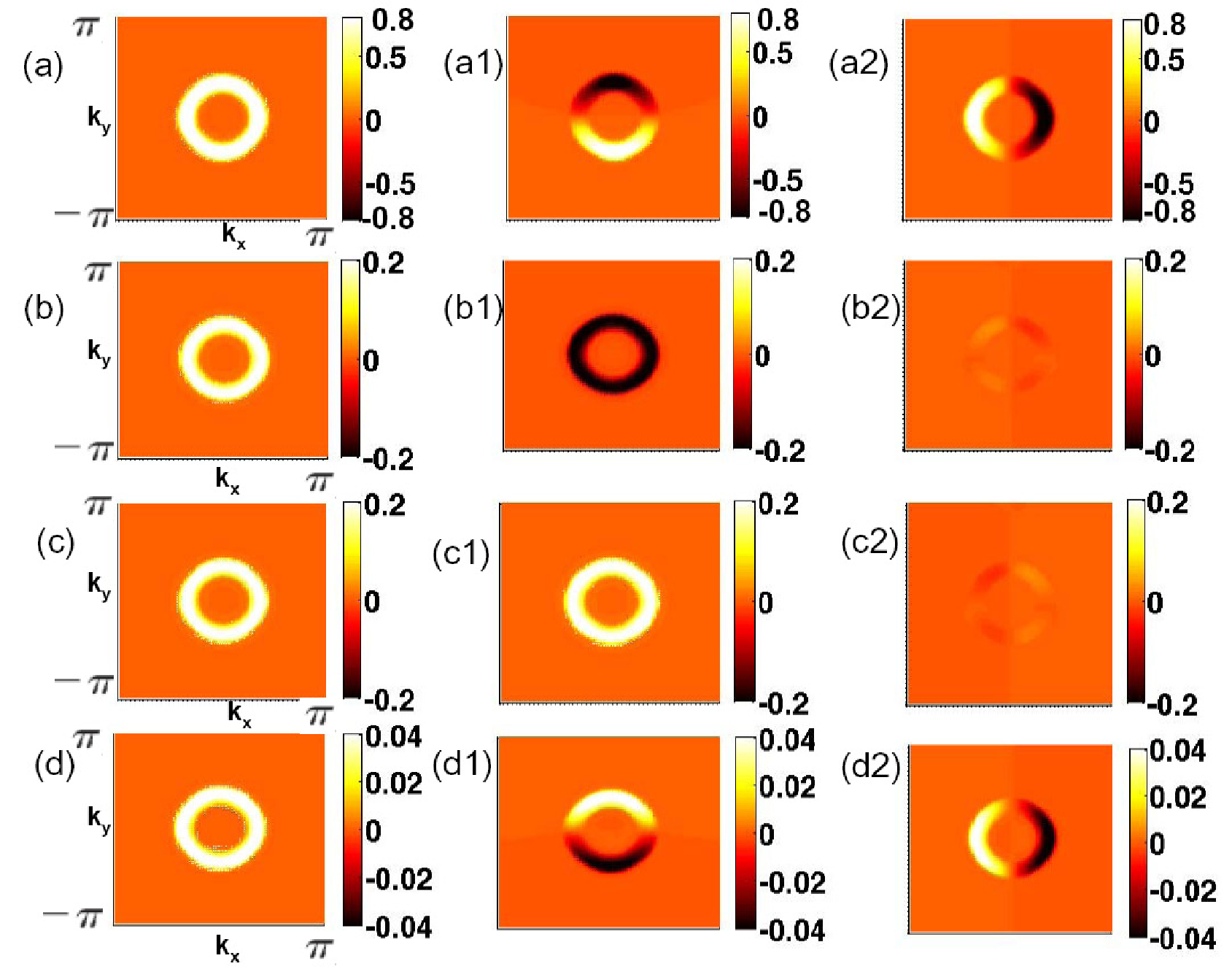}
\end{center}
\caption{(Color online) The self-consistent results of EOPs versus momentum
($k_{x},k_{y}$). (a), (b), (c) and (d) refer to $\left\vert \Delta
_{\uparrow\uparrow}(\mathbf{k})\right\vert $, $\left\vert \Delta
_{\uparrow\uparrow}(\mathbf{k})\right\vert $, $\left\vert \Delta
_{\downarrow\uparrow}(\mathbf{k})\right\vert $ and $\left\vert \Delta
_{\downarrow\downarrow}(\mathbf{k})\right\vert .$Correspondingly, (a1), (b1),
(c1) and (d1) refer to Re($\Delta_{\uparrow\uparrow}(\mathbf{k})$),
Re($\Delta_{\uparrow\uparrow}(\mathbf{k})$), Re($\Delta_{\downarrow\uparrow
}(\mathbf{k})$) and Re($\Delta_{\downarrow\downarrow}(\mathbf{k})$); (a2),
(b2), (c2) and (d2) refer to Im($\Delta_{\uparrow\uparrow}(\mathbf{k})$),
Im($\Delta_{\uparrow\uparrow}(\mathbf{k})$), Im($\Delta_{\downarrow\uparrow
}(\mathbf{k})$) and Im($\Delta_{\downarrow\downarrow}(\mathbf{k})$). The
effective Zeeman fields are ($V_{e}$, $V_{h}$)=($-1$, $-1$).}%
\end{figure}

\begin{figure}[ptb]
\begin{center}
\includegraphics[width=1.0\linewidth]{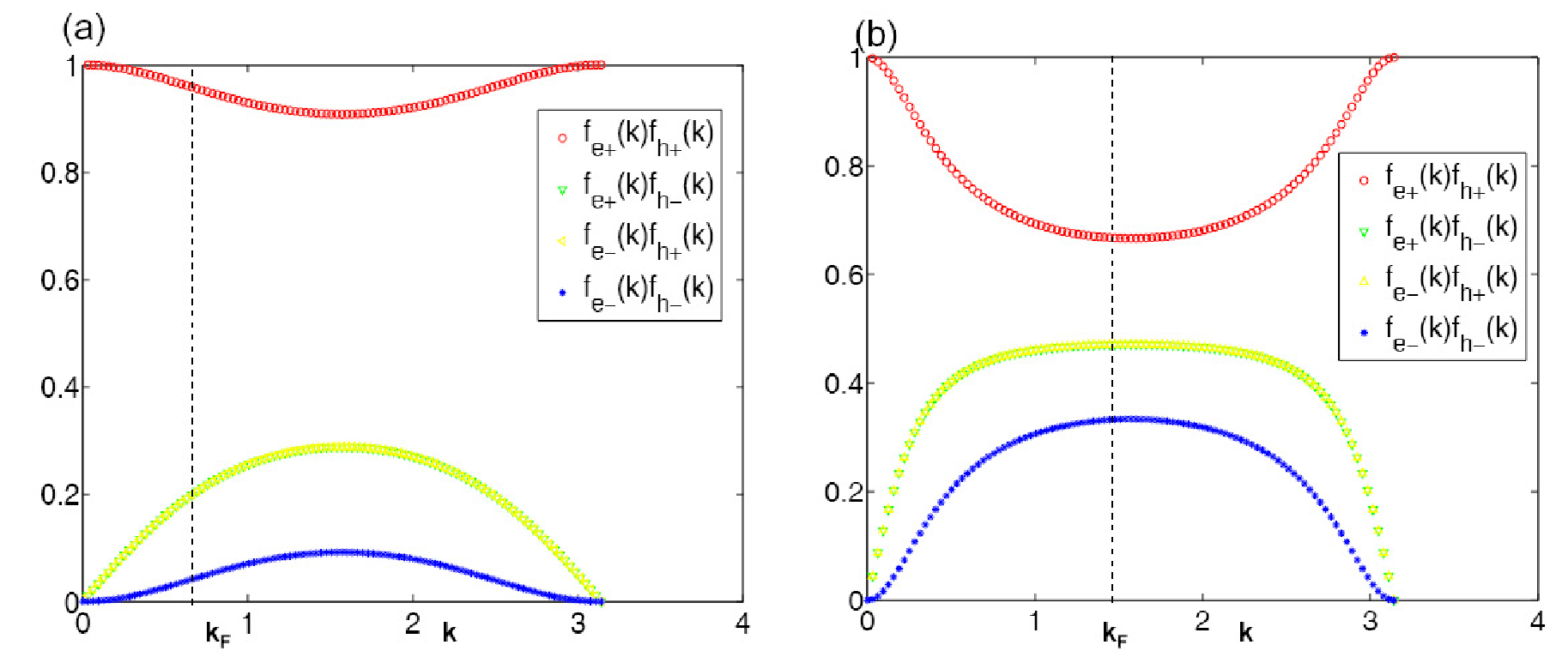}
\end{center}
\caption{(Color online) The coefficients $f_{e+}(\mathbf{k})f_{h+}%
(\mathbf{k})$, $f_{e+}(\mathbf{k})f_{h-}(\mathbf{k})$, $f_{e-}(\mathbf{k}%
)f_{h+}(\mathbf{k})$ and $f_{e-}(\mathbf{k})f_{h-}(\mathbf{k})$ versus the
momentum $k$ in the magnetization configuration ($V_{e},V_{h}$)=($-1,-1$) for
different Rashba SOC values: (a) $\alpha$=$0.5$; (b) $\alpha$=$2$. }%
\end{figure}

The factors $f_{e\pm}(\mathbf{k})f_{h\pm}(\mathbf{k})$ in Eq. (\ref{relation})
can be understood as follows. Since the four pairing states with different
spin combinations in EOPs are affected differently by the effective exchange
fields. Additionally, the Rashba SOC flaws the spin polarization of the
carries along the $z$ direction, then the total effect leads the factors
$f_{e\pm}(\mathbf{k})f_{h\pm}(\mathbf{k})$ to emerge in four pairing states of
EOPs and determine the EOP behavior at given ($V_{e},V_{h},$). We find that
only one component of $\Delta_{\sigma\sigma^{\prime}}(\mathbf{k})$ is
dominant, while the other components can be neglected in the proper range of
the Rashba SOC parameter for the given magnetization of the FM films.
Moreover, the system's properties are decided by the dominant component of
$\Delta_{\sigma\sigma^{\prime}}(\mathbf{k})$. For definiteness, we have
carried out a series of self-consistent numerical solutions of the BdG
equation (3) by setting the lattice size 81$\times$81, $\mu_{e}$=$\mu_{h}%
$=$-4$, $\alpha$=0.5, and ($V_{e},V_{h}$)=($\pm1,\pm1$). As a typical example,
we show in Fig. 2 the numerical results of $\Delta_{\sigma\sigma^{\prime}%
}(\mathbf{k})$ for ($V_{e},V_{h}$)=($-1,-1$). Also, we present in Fig. 3(a)
the results of the factors $f_{e\pm}(\mathbf{k})f_{h\pm}(\mathbf{k})$ for
comparison. In this case, from Fig. 2 it is clear that $\Delta_{\uparrow
\uparrow}(\mathbf{k})$ is the dominant EOP component which has the phase of
$-ie^{-i\phi_{\mathbf{k}}}$, i.e. $e^{i(\theta_{\mathbf{k}}-\vartheta
_{\mathbf{k}})}$=1.0. In the continuous limit, $-ie^{-i\phi_{\mathbf{k}}}%
\sim-\frac{k_{y}+ik_{x}}{k}$, which means the $p$+$ip$-like pairing state.
Furthermore, this kind of pairing state is usually topologically nontrivial
(see discussion below). Similarly, for the case of ($V_{e},V_{h}$)=($1,1$),
the $\Delta_{\downarrow\downarrow}(\mathbf{k})$ turns out to be the dominant
pairing component with a phase of $-ie^{i\phi_{\mathbf{k}}}$, which thus is
also topologically nontrivial. On the other hand, for the case of
($V_{e},V_{h}$)=($-1,1$), the $\Delta_{\uparrow\downarrow}(\mathbf{k})$ is the
dominant pairing component with a phase of $-$($e^{i\phi_{\mathbf{k}}}%
$+$e^{-i\phi_{\mathbf{k}}}$). The physics related to this kind of phase is
that $e^{i(\theta_{\mathbf{k}}-\vartheta_{\mathbf{k}})}$=$e^{\pm
i\phi_{\mathbf{k}}}$ correspond to two degenerate states and their linear
combinations give the ground state with the phase of $-$($e^{i\phi
_{\mathbf{k}}}$+$e^{-i\phi_{\mathbf{k}}}$). In the continuum limit,
$-$($e^{i\phi_{\mathbf{k}}}$+$e^{-i\phi_{\mathbf{k}}}$) $\sim-\frac{k_{x}}{k}%
$. This kind of pairing state is topologically trivial. Similarly, for the
case of ($V_{e},V_{h}$)=($1,-1$), the dominant pairing component
$\Delta_{\downarrow\uparrow}(\mathbf{k})$ is also topologically trivial.

Furthermore, from Fig. 3 (a) and Eq. (7) we can find that the weak Rashba SOC
relative to the Zeeman fields will select one dominating component of EOPs
(namely, $\Delta_{\uparrow\uparrow}(\mathbf{k})$), while other components can
be neglected. But in Fig. 3 (b) we can find that the amplitudes of
$f_{e+}(k)f_{h+}(k)$, $f_{e+}(k)f_{h-}(k)$ and $f_{e-}(k)f_{h+}(k)$ are
comparable. That means the non-weak Rashba SOC relative to the exchange field
cannot select one dominating component of EOPs, and all components of EOPs
play important roles. The consequence is that the unconventional pairing
states become unstable in case of Fig. 3 (b), where the value of $\alpha$ is
chosen to be much larger than the exchange filed. We can also clarify it from
numerical calculation. Hence, the Rashba SOC cannot be larger than the Zeeman
splitting in our system. Fortunately, it is easy to satisfy this requirement,
because both Rashba SOC and Zeeman fields are tunable in our system. Besides,
another essential point we stress is that the topologically trivial or
nontrivial EC in our system is stable, which means a stable bulk gap
separating the ground state and the excited states. We can obviously find this
fact in Fig. 1(b) and Fig. 4(a) (see below). In the following section, we
focus on the characterization of the topological properties of the system.

\section{ Chiral topological order}

From the self-consistent results of the aforementioned section, we can find
that the EOPs are closely correlated to the magnetization configuration of the
exchange fields. We also note that there are two different classes:
($V_{e},V_{h}$)=($\pm1,\pm1$) and ($V_{e},V_{h}$)=($\pm1,\mp1$). So we choose
one configuration from each class as an example. The EOPs can be simplified as
$\Delta_{\uparrow\uparrow}(\mathbf{k})$=$-i\Delta_{0}(\sin k_{x}-i\sin k_{y})$
and other $\Delta_{\sigma\sigma^{\prime}}(\mathbf{k})\mathtt{\sim}$0 for
($V_{e},V_{h}$)=($-1,-1$), while $\Delta_{\uparrow\downarrow}(\mathbf{k}%
)$=$-\Delta_{0}\sin k_{x}$ and other $\Delta_{\sigma\sigma^{\prime}%
}(\mathbf{k})\mathtt{\sim}$0 for ($V_{e},V_{h}$)=($-1,1$). Then the mean-field
Hamiltonian Eq. (\ref{Ham1_mf1}) in these two magnetization configurations can
be adiabatically mapped into the following forms, which are topologically
equivalent to initial mean-field Hamiltonian Eq. (\ref{Ham1_mf1}) when
$\alpha\rightarrow0$ without the bulk gap closing \cite{Sato1}.
\begin{equation}
\mathcal{H}_{MF}^{t}=\sum_{\mathbf{k}}[\hat{e}_{\mathbf{k\uparrow}},\hat
{h}_{-\mathbf{k\uparrow}}^{\dagger}]H^{t}(\mathbf{k})\left[
\begin{array}
[c]{c}%
\hat{e}_{\mathbf{k\uparrow}}^{\dagger}\\
\hat{h}_{-\mathbf{k\uparrow}}%
\end{array}
\right]  \label{H_t}%
\end{equation}
$,$%
\begin{equation}
\mathcal{H}_{MF}^{s}=\sum_{\mathbf{k}}[\hat{e}_{\mathbf{k\uparrow}},\hat
{h}_{-\mathbf{k\downarrow}}^{\dagger}]H^{s}(\mathbf{k})\left[
\begin{array}
[c]{c}%
\hat{e}_{\mathbf{k\uparrow}}^{\dagger}\\
\hat{h}_{-\mathbf{k\downarrow}}%
\end{array}
\right]  \label{H_s}%
\end{equation}
where $H^{t}(\mathbf{k})=(\zeta_{\mathbf{k}}^{(e)}-\mu_{e}+V_{e})(s_{0}%
+s_{3})/2+(-\zeta_{\mathbf{k}}^{(h)}+\mu_{h}-V_{h})(s_{0}-s_{3})/2-i\Delta
_{0}(\sin k_{x}s_{1}+\sin k_{y}s_{2})$ and $H^{s}(\mathbf{k})=(\zeta
_{\mathbf{k}}^{(e)}-\mu_{e}+V_{e})(s_{0}+s_{3})/2+(-\zeta_{\mathbf{k}}%
^{(h)}+\mu_{h}+V_{h})(s_{0}-s_{3})/2-\Delta_{0}\sin k_{x}s_{1}.$ It is easy to
check that both $H^{s}(\mathbf{k})$ and $H^{t}(\mathbf{k})$ preserve the PHS
and break the TRS according to the definition of Ref. \cite{Schnyder}. The
missing of TRS can be traced back to the effective Zeeman fields in Eq.
(\ref{Ham1_mf1}). From Ref. \cite{Schnyder}, we can find that the Hamiltonian
describing our system generally belongs to class $D$ in BdG class. The
topologically nontrivial pairing has the form of chiral $p$ wave (namely,
$p_{x}+ip_{y}$ or $p_{x}-ip_{y}$ pairing) in two dimensions. The $p_{x}$ or
$p_{y}$ pairing is trivial. Furthermore, the nontrivial topology is
characterized by an integer which is similar to the TKNN number in the QHE.

We can calculate this topological number from Eq. (\ref{Ham1_mf1}). The
topological nature of the ground state of the exciton condensate can be
characterized by non-zero $I_{TKNN}$, which reads%
\begin{equation}
I_{TKNN}=-\frac{1}{2\pi}\sum\limits_{n=1}^{2}\int_{BZ}d^{2}k\Omega
_{n}(\mathbf{k}), \label{chern}%
\end{equation}
where $\Omega_{n}(\mathbf{k})$=$-2\operatorname{Im}\left\langle \frac{\partial
u_{n}(\mathbf{k})}{\partial k_{x}}\right\vert \left.  \frac{\partial
u_{n}(\mathbf{k})}{\partial k_{y}}\right\rangle $ is the Berry curvature of
the occupied energy band $n$ and$\left\vert u_{n}(\mathbf{k})\right\rangle $
is the corresponding Bloch wave function. The straightforward calculation
gives $I_{TKNN}$=$1$ for ($V_{e},V_{h}$)=($-1,-1$) and $I_{TKNN}$=$0$ for
($V_{e},V_{h}$)=($-1,1$). From the bulk-edge correspondence, the nontrivial
bulk topological number implies gapless edge states emerging in the terminal
of the system.

\begin{figure}[ptb]
\begin{center}
\includegraphics[width=1.0\linewidth]{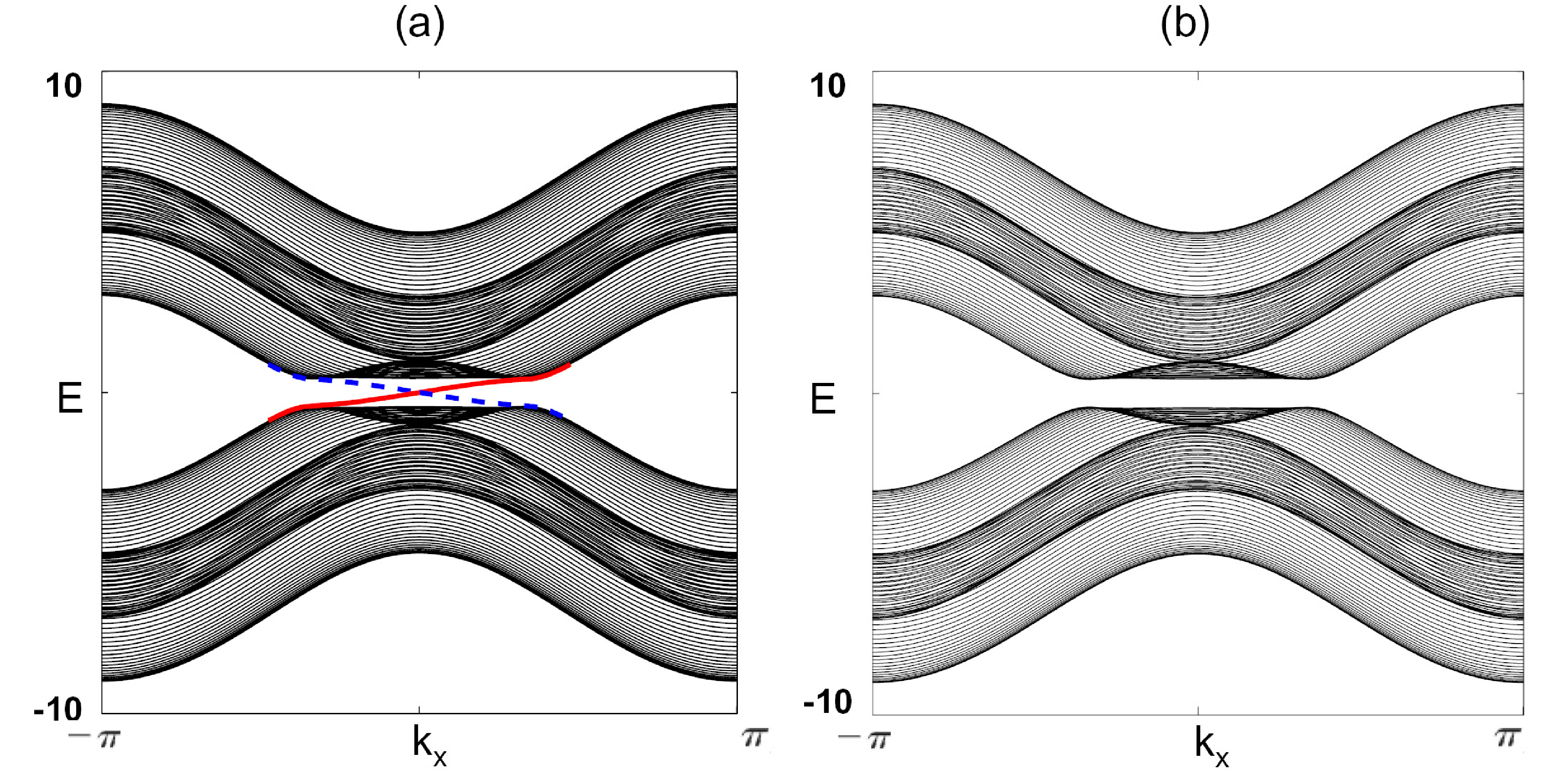}
\end{center}
\caption{(Color online) The energy spectrum of the bilayer square-lattice
system with two edges at the $y$ direction. $k_{x}$ denotes the momentum in
the $x$ direction. The magnetization parameters are set at ($V_{e},V_{h}%
$)=($-1,-1$) for (a) and ($V_{e},V_{h}$)=($-1,1$) for (b). In both cases
$\Delta_{0}$=0.5 and $\alpha$=0.5. The red-solid and blue-dashed lines denote
the edge states locating at different edges.}%
\end{figure}

In order to evaluate gapless edge states, we assume that the square-lattice
system has two edges in $y$ direction and is boundless in $x$ direction.
Correspondingly, we choose open boundary condition in $y$ direction and
periodic boundary condition in $x$ direction in the lattice Hamiltonian in Eq.
(\ref{H_lat}) in mean-field approximation. The calculated energy spectrums of
the topologically nontrivial cases of ($V_{e},V_{h}$)=($-1,-1$) and
topologically trivial case of ($V_{e},V_{h}$)=($-1,1$) are illustrated in Fig.
4 (a) and (b), respectively, in which the red-solid and blue-dashed lines
correspond to the different edge states with opposite chiralities. It is
obvious to find that the number of the gapless edge states is consistent with
the bulk theory characterized by $I_{TKNN}$.

Furthermore, for the case of ($V_{e},V_{h}$)=($-1,-1$), the edge states can be
described by the effective Hamiltonian:%

\begin{equation}
H_{edge}=\pm\underset{k_{x}\geq0}{\sum}v_{F}k_{x}\hat{\gamma}^{\dagger}%
(k_{x},y)\hat{\gamma}(k_{x},y),\label{edge Ham}%
\end{equation}
where $\pm$ represents the opposite chiralities, $v_{F}$ is the Fermi
velocity, and $k_{x}$ is the momentum measured from the Fermi surface, and
\begin{equation}
\hat{\gamma}(k_{x},y)=u(k_{x},y)\hat{e}_{\uparrow}(y)+v(k_{x},y)\hat
{h}_{\uparrow}^{\dagger}(y).\label{quasi oper}%
\end{equation}
We will show in next section that the low-energy topological quasiparticle
excitations are dominated by the edge states.

\section{Fractional qusiparticle excitations in Exciton Condensate}

We now demonstrate that there exist fractional excitations in our system in
the topologically nontrivial phases if the EOPs have a $U(1)$ vortex profile
with odd vorticity. Note that the even vorticity is trivial \cite{Gauarie}.
For definiteness, we consider the case: ($V_{e},V_{h}$)=($-V,-V$)=($-1,-1$)
and $\mu_{e}$=$\mu_{h}$=$\mu$=$-4t_{p}$. The non-vortex BdG Hamiltonian in Eq.
(\ref{Ham1_mf1}) reads%

\begin{equation}
\mathcal{H}_{MF}(\mathbf{k})=\left[
\begin{array}
[c]{cc}%
\mathbf{\Sigma}_{\mathbf{k}}^{(e)}-\mu-Vs_{z} & \mathbf{\Delta}(\mathbf{k})\\
\mathbf{\Delta}^{\dagger}(\mathbf{k}) & \mathbf{\Sigma}_{-\mathbf{k}}%
^{(h)}+\mu+Vs_{z}%
\end{array}
\right]  \label{H_BdG}%
\end{equation}
with%

\begin{equation}
\mathbf{\Delta}(\mathbf{k})=\left[
\begin{array}
[c]{cc}%
-i\Delta_{0}(\sin k_{x}-i\sin k_{y}) & 0\\
0 & 0
\end{array}
\right]  . \label{EOP_BdG}%
\end{equation}
The PHS of $\mathcal{H}_{MF}(\mathbf{k})$ is defined by $\Lambda
\mathcal{H}_{MF}(\mathbf{k})\Lambda$=$-\mathcal{H}_{MF}(-\mathbf{k})$ with
$\Lambda$=$s_{x}\otimes s_{0}\mathcal{K}$, where $\mathcal{K}$ is the
complex-conjugation operator and $\Lambda^{2}$=$1$.

Now, we consider a vortex in the EOPs, i.e., the uniform $\Delta_{0}$ in Eq.
(14) is modulated to\texttt{ }$f^{2}(r)e^{in\theta}$, where $r$ and $\theta$
are polar coordinates centered on the vortex, $n$ is the vorticity, and $f(r)$
is a real function of $r$ that vanishes at small $r$ and obtains asymptotic
value $f_{0}$ (a real constant) at large $r$. Usually, the nature of the
excitation bound to the vortex is dominated by the low-energy limit of the
system, i.e. $\mathbf{k\mathtt{\rightarrow}}0$. In this limit, the effective
Hamiltonian can be obtained by the replacement $\mathbf{k}$
$\mathtt{\rightarrow}-i\mathbf{\nabla}$ and is accurate up to $\mathcal{O}%
(\mathbf{k}^{2}).$ The zero-mode equation reads $\mathcal{H}_{MF}%
(-i\mathbf{\nabla})\Psi_{\mathbf{0}}$=$0$ with $\Psi_{\mathbf{0}}(r,\theta
)$=[$u_{\uparrow},u_{\downarrow},v_{\uparrow},v_{\downarrow}$]$^{T}$. With the
help of $\Lambda\mathcal{H}_{MF}(-i\mathbf{\nabla})\Lambda$=$-\mathcal{H}%
_{MF}(-i\mathbf{\nabla})$, it is easy to find $v_{\sigma}$=$u_{\sigma}^{\ast}%
$. Hence the zero-mode equations can be expressed as%

\begin{align}
-(\frac{\partial^{2}}{\partial z\partial z^{\ast}}+V)u_{\uparrow}+\alpha
\frac{\partial u_{\downarrow}}{\partial z^{\ast}}-f(r)e^{in\theta/2}%
\frac{\partial(f(r)e^{in\theta/2}v_{\uparrow})}{\partial z^{\ast}}  &
=0,\label{Eq_zero mode a}\\
-\alpha\frac{\partial u_{\uparrow}}{\partial z}+(-\frac{\partial^{2}}{\partial
z\partial z^{\ast}}+V)u_{\downarrow}  &  =0, \label{Eq_zero mode b}%
\end{align}
where $z$=$x$+$iy$. From Eq. (\ref{quasi oper}) in Sec. IV, we can find that
$u_{\downarrow}$ can be neglected. Moreover, the numerical results shown in
Fig. 5 also give a proof. Hence, we only need to consider Eq.
(\ref{Eq_zero mode a}). For definiteness we consider the simplest case in the
quantum limit analogous to the London approximation in superconductor, i.e.,
$f(r$=$0)$=$0$, $f(r$%
$>$%
$0)$=$f_{0}$ (a positive real constant) and the $U(1)$ vortex with vorticity
$n$=1. Moreover, we look for a solution in terms of a spherically symmetric
real function $u_{\uparrow}$. Note that in Eq. (\ref{Eq_zero mode b}), we must
take into account the $U(1)$ symmetry in order to get the physical solutions.
Then Eq. (\ref{Eq_zero mode a}) can be simplified into the following
differential equation:%

\begin{equation}
\frac{d^{2}u_{\uparrow}}{dr^{2}}+(\frac{1}{r}+f_{0}^{2})\frac{du_{\uparrow}%
}{dr}+(\frac{f_{0}^{2}}{2r}+V)u_{\uparrow}=0\text{.} \label{Eq_zero mode 2}%
\end{equation}
\begin{figure}[ptb]
\begin{center}
\includegraphics[width=1.0\linewidth]{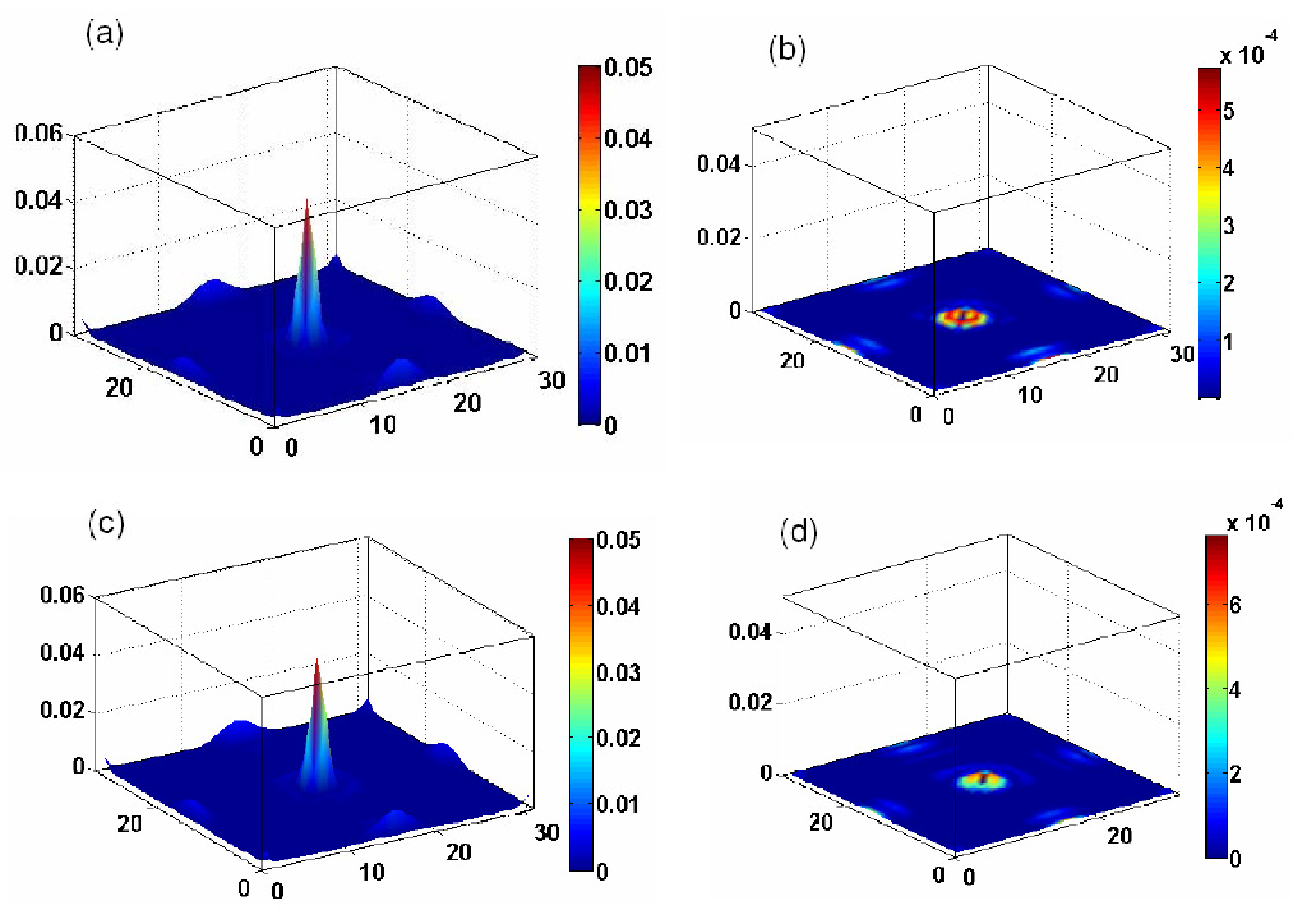}
\end{center}
\caption{(Color online) The density distributions of quasiparticle excitation
of the $U(1)$ vortex locating on the center of a $31\times31$ lattice for (a)
$\left\vert u_{\uparrow}(r)\right\vert ^{2}$, (b) $\left\vert u_{\downarrow
}(r)\right\vert ^{2}$, (c) $\left\vert v_{\uparrow}(r)\right\vert ^{2}$, and
(d) $\left\vert v_{\downarrow}(r)\right\vert ^{2}$ in the exchange fields
($V_{e},V_{h}$)=($-1,-1$). We take $f_{0}^{2}$=0.5, $\alpha$=0.5.}%
\end{figure}The solution of Eq. (\ref{Eq_zero mode 2}) is $u_{\uparrow}\propto
J_{0}(\sqrt{4V^{2}+f_{0}^{4}}r/2)e^{-f_{0}^{2}r/2}$ where $J_{0}(x)$ is the
Bessel function. We also give the exact numerical results of $\left\vert
u_{\sigma}\right\vert ^{2}$and $\left\vert v_{\sigma}\right\vert ^{2}$ in the
square-lattice system, which are shown in Fig. 5. Clearly, it can be seen that
$u_{\downarrow}\mathtt{\sim}0$, which is consistent with Eq. (\ref{quasi oper}).

The quasiparticle excitation field associated with the zero-mode solution in
the exciton condensate can be expressed as $\hat{\gamma}_{ec}$=$\int
d\mathbf{r}(u_{\uparrow}\hat{e}_{\uparrow}(\mathbf{r})+v_{\uparrow}\hat
{h}_{\uparrow}^{\dagger}(\mathbf{r}))$. Contrastively, the quasiparticle
excitation field describing the vortex zero mode in topological superconductor
is $\hat{\gamma}_{sc}$=$\int d\mathbf{r}(u_{\uparrow}\hat{c}_{\uparrow
}(\mathbf{r})+v_{\uparrow}\hat{c}_{\uparrow}^{\dagger}(\mathbf{r}))$
\cite{Sato,Qi}. It is obvious to find that $\hat{\gamma}_{ec}\neq\hat{\gamma
}_{ec}^{\dagger}$ due to $[\hat{e}_{\uparrow}(\mathbf{r}),\hat{h}_{\uparrow
}^{\dagger}(\mathbf{r})]_{+}$=$0$ and $\hat{\gamma}_{sc}=\hat{\gamma}%
_{sc}^{\dagger}$ due to $[\hat{c}_{\uparrow}(\mathbf{r}),\hat{c}_{\uparrow
}^{\dagger}(\mathbf{r})]_{+}$=$1$. That means the quasiparticle excitations in
topological superconductor are the Majorana fermions which obey the
non-Abelian anyonic statistics \cite{Read}. In our case, the quasiparticle
excitations are not the Majorana fermions and are charged. Generally, a
localized zero mode in a system with PHS is known to carry a fractional charge
of $\pm e/2$ \cite{Jackiw,Goldstone}. Based on the arguments given in Refs.
\cite{Herbut,Franz}, we expect for the present system that the quasiparticle
excitation bounding to vortice has fractional charge $e/2$. We can also
evaluate it from the spatial integration $\int d\mathbf{r}(\left\vert
u_{\uparrow}\right\vert ^{2}+\left\vert v_{\uparrow}\right\vert ^{2})$ around
the vortex defect. Fractional charged quasiparticles confined to 2D often obey
fractional exchange statistics \cite{Laughlin}, which is in our case Abelian
from the standard arguments \cite{Wilczek}.

\section{Conclusion}

In conclusion, we have studied the exciton condensate in zero temperature
limit in a special class of effective TCI bilayer system which is different
from the semiconductor bilayer system in Ref. \cite{Hao} due to its
preservation of PHS. With the self-consistent mean-field method, we have found
that the Rashba SOC in the electron and hole layers can induce the
unconventional $p\pm ip$ or $p$ pairing states that depend on the
magnetization of the FM insulating films. Correspondingly, the $p\pm ip$
pairing states represent topologically nontrivial exciton condensate phase,
which can be characterized by nonzero TKNN number or the gapless edge states,
while the $p$ paring states are topologically trivial. Furthermore, by
low-energy analytic solution as well as exact numerical calculation, we have
shown that in the topologically nontrivial exciton condensate phase, the
presence of $U(1)$ vortexes result in the simple Fermi zero-mode fractional
quasiparticle excitations that obey Abelian statistics, which completely
distinguish from the non-Abelian Majorana Fermi zero-mode quasiparticle
excitations in topological superconductor systems.

\begin{acknowledgments}
This work was supported by NSFC under Grants No. 90921003, No. 10574150, and
No. 60776063, and by the National Basic Research Program of China (973
Program) under Grants No. 2009CB929103 and No. 2011CB921701.
\end{acknowledgments}

\end{document}